\documentclass[twocolumn]{aastex63}

\usepackage{lineno}

\shorttitle{Magnetic field estimates using the MIT with absorption}

\usepackage{natbib}
\newcommand{\fex}{\ion{Fe}{10}}

\newcommand{\longacknowledgment}{We gratefully acknowledge support by NASA grants 80NSSC18K1285, 80NSSC20K1272, 80NSSC21K0737, 80NSSC21K1684, and contract NNG09FA40C (IRIS). Enrico Landi gratefully acknowledges support by NASA grants 80NSSC20K0185 and 80NSSC22K0750. The simulations have been run on clusters the Pleiades cluster through the computing project s1061, s2601, and s8305 from the High End Computing (HEC) division of NASA. To analyze the data we have used IDL. This research is also supported by the Research Council of Norway through its Centres of Excellence scheme, project number 262622, and through grants of computing time from the Programme for Supercomputing. The synthesis and analysis of the numerical models have been performed with the aid of Google Cloud Platform (GCP, https://console.cloud.google.com, project  lunar-campaign-29341), allowing sharing of the synthetic data and models, developing common tools, and access to instances with various specifications and Graphics Processing Units (GPUs).}

\begin{document}

\title{A novel inversion method to determine the coronal magnetic field including the impact of bound-free absorption}

\author[0000-0002-0333-5717]{Juan Mart\'inez-Sykora}
\affil{Lockheed Martin Solar \& Astrophysics Laboratory,
3251 Hanover St, Palo Alto, CA 94304, USA}
\affil{Bay Area Environmental Research Institute,
NASA Research Park, Moffett Field, CA 94035, USA.}
\affil{Rosseland Center for Solar Physics, University of Oslo, P.O. Box 1029 Blindern, N-0315 Oslo, Norway}
\affil{Institute of Theoretical Astrophysics, University of Oslo,
P.O. Box 1029 Blindern, N-0315 Oslo, Norway}

\author[0000-0003-0975-6659]{Viggo H. Hansteen}
\affil{Lockheed Martin Solar \& Astrophysics Laboratory,
3251 Hanover St, Palo Alto, CA 94304, USA}
\affil{Bay Area Environmental Research Institute,
NASA Research Park, Moffett Field, CA 94035, USA.}
\affil{Rosseland Center for Solar Physics, University of Oslo, P.O. Box 1029 Blindern, N-0315 Oslo, Norway}
\affil{Institute of Theoretical Astrophysics, University of Oslo,
P.O. Box 1029 Blindern, N-0315 Oslo, Norway}

\author[0000-0002-8370-952X]{Bart De Pontieu}
\affil{Lockheed Martin Solar \& Astrophysics Laboratory,
3251 Hanover St, Palo Alto, CA 94304, USA}
\affil{Rosseland Center for Solar Physics, University of Oslo, P.O. Box 1029 Blindern, N-0315 Oslo, Norway}
\affil{Institute of Theoretical Astrophysics, University of Oslo,
P.O. Box 1029 Blindern, N-0315 Oslo, Norway}

\author[0000-0002-9325-9884]{Enrico Landi}
\affil{Department of Climate and Space Sciences and Engineering, University of Michigan, Ann Arbor, MI, USA}

\begin{abstract}
The magnetic field governs the corona; hence it is a crucial parameter to measure. Unfortunately, existing techniques for estimating its strength are limited by strong assumptions and limitations. These techniques include  photospheric or chromospheric field extrapolation using potential or non-linear-force-free methods, estimates based on coronal seismology, or by direct observations via, e.g., the Cryo-NIRSP instrument on DKIST which will measure the coronal magnetic field, but only off the limb. Alternately, in this work we investigate a recently developed approach based on the magnetic-field-induced (MIT) transition of the \fex~257.261~\AA. In order to examine this approach, we have synthesized several \fex\ lines from two 3D magnetohydrodynamic simulations, one modeling an emerging flux region and the second an established mature active region. In addition, we take bound-free absorption from neutral hydrogen and helium and singly ionised helium into account. The absorption from cool plasma that occurs at coronal heights has a significant impact on determining the magnetic field. We investigate in detail the challenges of using these \fex\ lines to measure the field, considering their density and temperature dependence. We present a novel approach to deriving the magnetic field from the MIT using inversions of the differential emission measure as a function of the temperature, density, and magnetic field. This approach successfully estimates the magnetic field strength (up to \%18 relative error) in regions that do not suffer from significant absorption and that have relatively strong coronal magnetic fields ($>250$~G). This method allows the masking of regions where absorption is significant.  
\end{abstract}


\section{Introduction} \label{sec:intro}

The evolution and strength of the solar coronal magnetic field is of great importance in understanding the driving of energetic flares \citep{Benz:2017LRSP...14....2B} and coronal mass ejections (CMEs) \citep{Chen:2011LRSP....8....1C}. It forms the framework upon which the energy transport in coronal waves occurs \citep{Cranmer:2005ApJS..156..265C}, and furthermore forms complex structures and arcades in active regions \citep{Schrijver:1997xu}. In the quiet sun, the magnetic field connects the corona with photospheric motions, mainly through the chromospheric network where the energetics are dominated by phenomena such as shock waves, spicules, and other types of jets \citep{Martinez-Sykora:2017sci,DePontieu2021}. Moreover, the magnetic field links the layers from the solar interior, where the field is created, to the outer bounds of our heliosphere more than 100~AU from the Sun. Beyond solar physics, the ability of measuring stellar fields is also crucial, for instance, in understanding superflares \citep{Maehara:2012Natur.485..478M}, or the impact of stellar CMEs \citep{Argiroffi:2019NatAs...3..742A} on exoplanets \citep{Dong:2017ApJ...837L..26D}. 

The magnetic field is measured at photospheric heights via the Zeeman and Hanle effects \citet{delToroIniesta:2016LRSP...13....4D,Asensio:2008ApJ...683..542A} using, for instance, spectropolarimeters \citep[eg., the Swedish Solar Telescope (SST), the Solar Optical Telescope on board of Hinede (SOT), or the Daniel K. Inouye Solar Telescope (DKIST)][]{Scharmer:2003ve, Tsuneta:2008kc,deWijn:2022SoPh..297...22D}. Recently, this has also been applied for strong non-LTE lines formed in the chromosphere, by inverting their full Stokes profiles \citep{delaCruzRodriguez:2019A&A...623A..74D}. However, Zeeman splitting is generally very weak and hence negligible for coronal lines. DKIST will have sufficient signal and spectral resolution in the infrared using the Cryo-NIRSP instrument  to study the coronal magnetic field \citep{Rast2021SoPh..296...70R}. However, such observations will be possible only at the solar off-limb where confusion along the line of sight (LOS) is possible and perhaps likely. Another approach to estimating the coronal magnetic field strength and topology lies in using photospheric (or chromospheric) magnetic measurement and performing field extrapolations based on these. Non-linear force free field (NLFFF) extrapolations are commonly used in the literature, but these have clear limitations since the photosphere and chromosphere are pressure gradient dominated \citep{DeRosa2015ApJ...811..107D} thus making the assumption of force free fields at least partially invalid. These methods can be aided or improved by the concurrent use of intensity maps from the transition region or coronal spectral lines \citep{Aschwanden:2016qy}. Similarly, the non force free nature of lower atmosphere fields plagues the more advanced magneto-friction numerical methods, which aspire to capture the time-dependent evolution of the magnetic field \citep[e.g.,][]{Cheung2012ApJ...757..147C}. Clearly, direct measurements of the coronal magnetic field would be an important supplement to these techniques.

Recent work suggests the use of magnetic-field-induced transition (MIT), e.g., the \fex~257.261~\AA\ line, to directly measure the coronal magnetic field \citep{Li:2015ApJ...807...69L,Li:2016ApJ...826..219L,Chen:2021ApJ...920..116C}. Such MIT transitions are highly sensitive to the strength of the magnetic field. However, the MIT transition is an unresolvable blend with two other \fex\ lines, which decreases the sensitivity of the total 257~\AA\ line to the magnetic field. Several methods have been proposed to measure the magnetic field using the the 257~\AA\ blended lines: originally, \citet{Si:2020ApJ...898L..34S} proposed line intensity ratios between the total intensity of the 257~\AA\ line with other \fex\ lines; later \citet{Landi:2020ApJ...904...87L} proposed a combination of intensity ratios capable of removing the blending contributions and improve the accuracy of the magnetic field measurement. However, regardless of the method, the intensity of the 257~\AA\ line is also dependent on the electron temperature and density, so that these two parameters need to be constrained in order to achieve a good estimate of the field strength. In addition, all MIT studies have ignored the absorption that the observed spectral line emission will suffer due to the presence of neutral hydrogen, helium and singly ionized helium.  \citet{Berger:1999ApJ...519L..97B,dePontieu:1999SoPh..190..419D,Schrijver:1999SoPh..187..261S,De-Pontieu:2003qr} pointed out that cool plasma may be present at coronal heights  (e.g., in the moss or upper transition region of hot loops) leading to EUV absorption. \citet{Anzer:2005ye} showed that EUV emission lines may expect significant bound-free absorption from neutral hydrogen, helium, and singly ionized helium and focused on such plasma in filaments and prominences. \citet{DePontieu:2009ApJ...702.1016D} used numerical models and observations to show evidence of this type of absorption in moss, while \citet{Hansteen:2019AA...626A..33H} showed it in the presence of recently emerged magnetic flux.

The MIT technique of measuring coronal magnetic fields has already been applied to existing observations. For example, \citet{Landi:2021fl} estimated the magnetic field during a C2 flare using Hinode/EIS. Similarly,  \citet{Brooks:2021ApJ...915L..24B} conducted a study of the field strength in coronal loops. \citet{Chen:2021ApJ...918L..13C} have tested the method for observations of stellar atmospheres. 

In this study, we will not only investigate the impact of absorption from high lying cool gas, but we will also present a new approach to inverting the observed data to determine the magnetic field by using the differential emission measure (DEM). This new approach is based on using the method of \citet{Cheung:2019ApJ...882...13C} to derive simultaneously the density, temperature and magnetic field of the coronal plasma. These inversions can also be used to disambiguate the spectrum from multi-slit observations, e.g., MUSE \citep{Cheung:2019ApJ...882...13C,DePontieu:2020ApJ...888....3D} or slitless spectrographs, e.g., COSIE \citep{Winebarger:2019ApJ...882...12W}. In the following, we give a short description of the forward radiative modeling code, that includes absorption (Section~\ref{sec:syn}). This code is used to compute the spectral line intensities in two different 3D radiative MHD models which are described in Section~\ref{sec:sim}. One is a simulation of an emerging flux region using the Bifrost code \citep[][]{Gudiksen:2011qy}. The results are discussed in the main text. The second simulation is of a mature active region \citep{Cheung:2019NatAs...3..160C} using MURaM \citep{Rempel:2017zl}, and results are summarized in the appendix. The results in Section \ref{sec:res} are composed of a first part that goes into the details of the properties of the various spectral lines with and without MIT and absorption (Section~\ref{sec:int}) and a second part where we apply the inversions to derive the magnetic fields and compare them to the models' ground truth (Section~\ref{sec:demb}). Section~\ref{sec:con} contains the conclusions and discussion. 

\section{Synthesis} \label{sec:syn}

The synthesized EUV spectral lines intensities ($I$) are computed assuming that the radiation is optically thin, and that the ionization state of the emitting ion is in statistical equilibrium with the temperature and density of the emitting plasma. However, we do allow for absorption by bound-free processes in overlying cold plasma: 

\begin{eqnarray}
I[\lambda] = \int_l n_e\, n_H\, G_\lambda (T, n_e, B)\, e^{-\tau}\,dl \label{eq:int}
\end{eqnarray}

\noindent where $n_e$, and $n_H$ are the electron and the hydrogen number densities, respectively. $G_\lambda(T, n_e, B)$ is the contribution function (see Appendix~\ref{sec:appendix_rad}).
The integration is carried out along the LOS ($l$), which is along the vertical axis mimicking disk-center observations for our models. In cases where we take the absorption into account, we have included absorption by neutral hydrogen, helium as well as singly ionized helium in computing $e^{-\tau}$ following the recipes of \citet{Anzer:2005ye} (see Appendix~\ref{sec:appendix_rad} for details).

This computation has been done using CHIANTI v.10 \citep{DelZanna:2021ApJ...909...38D} along with the Chiantipy software, and the integration has been done with GPUs using CUDA (pycuda): this allows us to use GPUs to vastly accelerate the calculations, which in turn allows the use of a finer grid along the LOS improving accuracy. We have assumed coronal abundances \citep{Feldman:1992ApJS...81..387F}. Further details on the model of \fex\ can be found in the Appendix~\ref{sec:appendix_rad}. 

Figure~\ref{fig:gtne} shows the contribution function of the lines of interest for this effort. All lines show some degree of density dependence, this is especially for \fex~174~\AA\ (panel a, e) and 257.261~\AA\ (panel c, g). It is also important to notice that the largest intensities are expected for the \fex~174~\AA\ line (panel a). Depending on the density, \fex~257.261~\AA\ (panel c) may have a much smaller contribution to the total intensity than the \fex~257.259~\AA\ line which it is blended with (panel b). 

\begin{figure*}
    \includegraphics[width=0.95\textwidth]{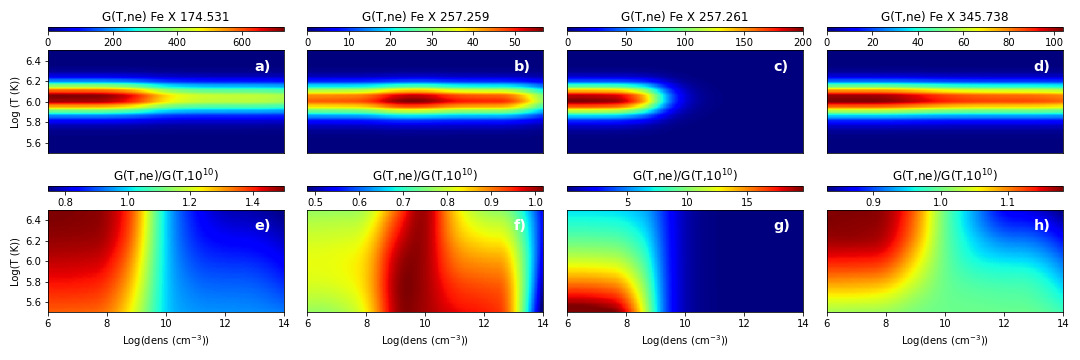}
	\caption{\label{fig:gtne} Contribution functions ($G(T,n_e,B)$) as a function of temperature and density for \fex\ 174, 257.259, 257.261 and 345~\AA\ are shown in the top row normalized by a factor of $10^{-27}$. The  contribution functions are normalized with $G(T)$ for a fixed density of $10^{10}$~cm$^{-3}$ in the bottom row. All lines have some degree of density dependence, but this is especially true for 174~\AA\ and 257.261~\AA. The plots for \fex~257.261~\AA~have zero magnetic fields ($B=0$~G).}
\end{figure*}

\begin{figure*}
    \includegraphics[width=0.95\textwidth]{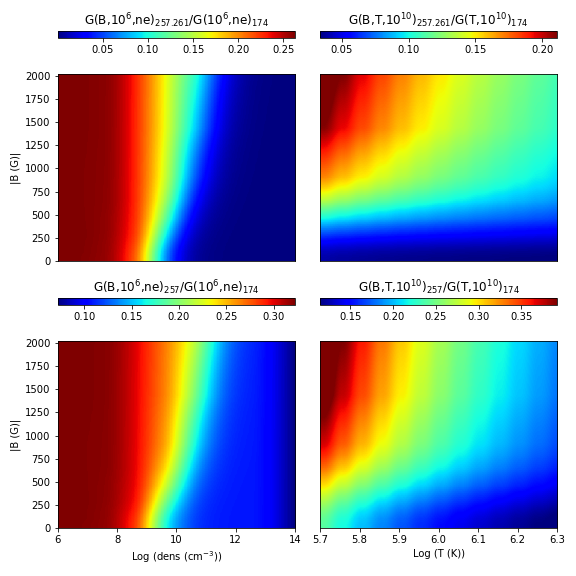}
	\caption{\label{fig:gbtne} Contribution function ($G(T, n_e, B)$) as a function of magnetic field and density (left) or temperature (right) for \fex~257.261~\AA\ (top) and the blended lines  \fex~257 (bottom). The contribution functions have been normalized with the contribution function of \fex~174~\AA .}
\end{figure*}

The magnetic field dependence of the contribution function for \fex\ 257.261~\AA\ ($G(T,n_e,B)_{257.261}$) is shown in Figure~\ref{fig:gbtne}, normalized with the contribution function of the 174~\AA\ line. Note that the magnetic field dependence varies strongly with density and temperature. The variation of the contribution function is greater in density and temperature than with magnetic field when comparing with $G(T,n_e)_{174}$. The bottom row shows the impact if including the \fex~257~\AA\ blend in the contribution functions: the ratio with the 174~\AA\ line is now less sensitive to the magnetic field. Note that we use photon units in contrast to \citet{Chen:2021ApJ...920..116C} who used energy units.  These properties between the various contribution functions need to be considered when used for density or magnetic field diagnostics. 

\section{Models} \label{sec:sim}

The current work uses a flux emergence simulation computed with the Bifrost code and an AR simulation computed with MURaM. The results from the MURaM simulation are collected and shown in the appendix. 

The Bifrost code \citep{Gudiksen:2011qy} is run with a plasma of solar photospheric abundance \citep{Asplund:2009uq} obeying an equation of state in thermodynamic equilibrium. Therefore, the temperature, pressure, ionization degree, and opacities at all points are computed (via a lookup table) from the density and internal energy. Radiative losses are given by optically thick radiation, including scattering, in four spectral bins, while radiative losses (and gains) in the upper chromosphere are treated according to  \cite{Carlsson:2012uq}. In the corona, optically thin losses are computed. In addition, thermal conduction along the magnetic field is included. To maintain an effective temperature close to that of the Sun, the entropy of inflowing gas is set at the bottom boundary, as is the strength and direction of the horizontal magnetic field as described below. The upper boundary in the corona uses characteristic methods to remain nearly transparent while the temperature gradient is set to zero. The horizontal boundaries are periodic.

The flux emergence simulation covers a computational domain of $72\times 72\times 60$~Mm$^3$, the vertical extent of which goes from the bottom boundary 8.5~Mm below the photosphere to the upper boundary 52~Mm above. This computational box is covered by $720\times720\times 1120$ grid points, giving a horizontal resolution of 100~km and a varying vertical simulation, with a 20~km resolution in the photosphere, chromosphere, and transition region stretching to 100~km in the deeper layers of the convection zone and the corona. The simulation starts with an initial horizontal field of 100~G up to the photosphere and close to 0~G in the overlying atmosphere. The simulation evolved for several hours with a horizontal magnetic field (a sheet) of 200~G injected at the bottom boundary for the first 95~minutes, which is ramped up to 1000~G for the next 70 minutes, then again to 2000~G for another 150~minutes. Thereafter a horizontal field of 300~G is injected continuously. The simulation has run for 8~hours solar time at the moment of the snapshot considered in this study. 

When magnetic flux becomes strong enough to pierce the photosphere and rise into the upper solar atmosphere it carries with it plasma of roughly photospheric temperatures. This cool gas drains back through the chromosphere to the photosphere eventually as the field continues to rise, but this can take significant time; thermal conduction is very inefficient perpendicular to the field preventing the plasma from heating to coronal temperatures, while at the same time the nearly horizontal field will prevent a rapid fall. Thus, at times when there is significant emergence of magnetic field, we expect, and find in the model, the corona to contain cool gas with temperatures of order $10^4$~K. This may be the case in emerging flux regions in the Sun, as found in the model used here. More generally, the total amount of cool gas found in the corona is not well known. A full description of this model and its evolution is in preparation \citep{Hansteen:2022prep}.

\begin{figure*}
    \includegraphics[width=0.5\textwidth]{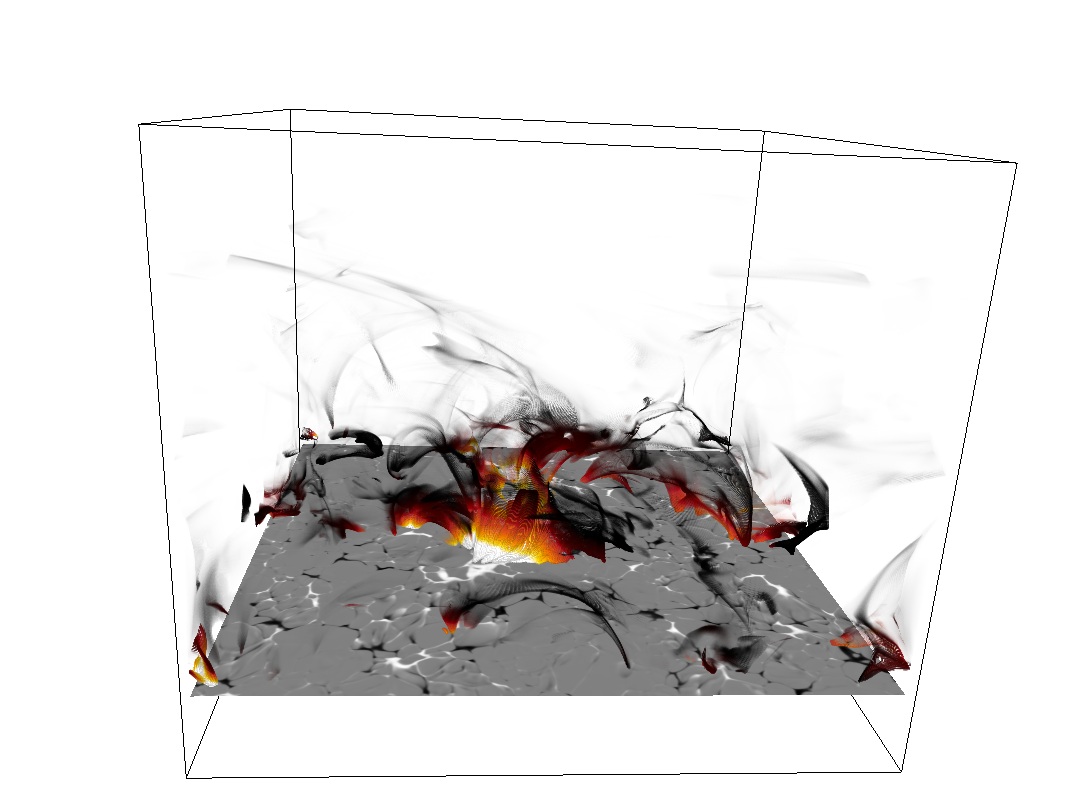}
    \includegraphics[width=0.5\textwidth]{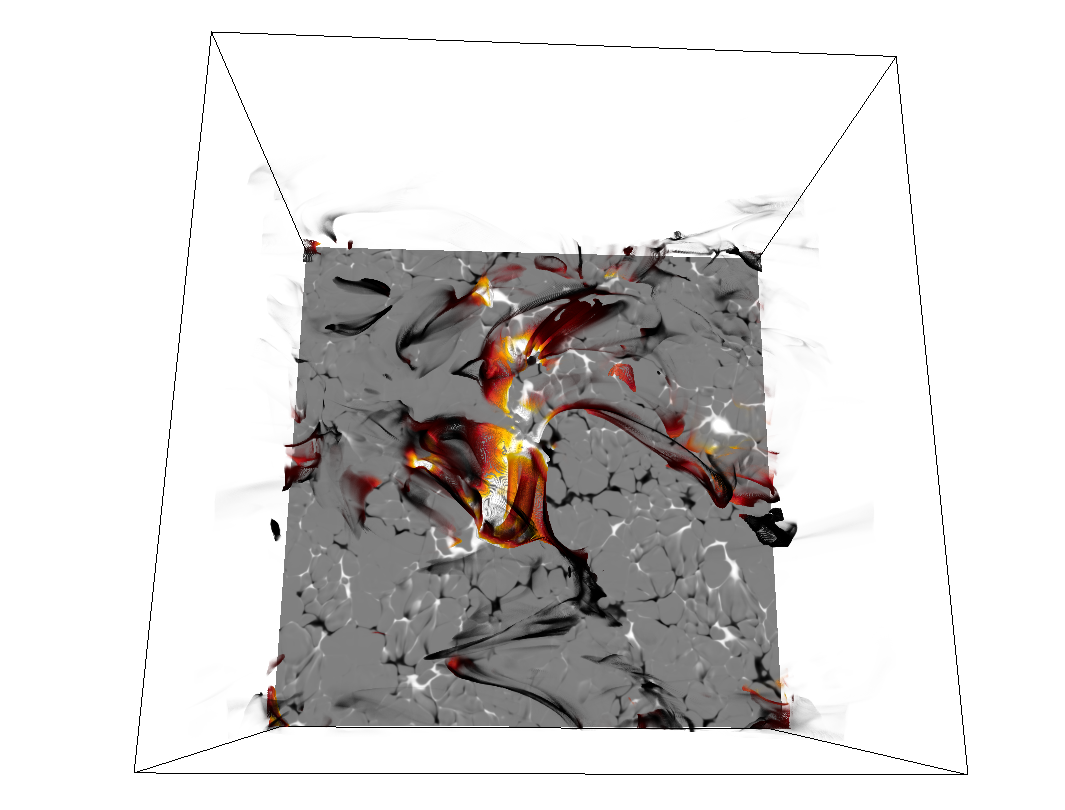}
	\caption{\label{fig:render} Three-dimensional rendering of the emissivity including bound-free absorption with increased rendering opacity for stronger emission. The color in the render from black through red to white represents the magnetic field strength, white being 300~G and black 50-G. The photospheric magnetic field is shown in the gray probe. }
	\label{fig:bifrost_field}
\end{figure*}

In Figure~\ref{fig:bifrost_field}, we show the vertical ($B_z$) component of the magnetic field in the photosphere along with the emissivity of the \fex~174.35 line, colored by the strength of the magnetic field. Low intensity regions are made transparent. The emissivity is multiplied by $\exp(-\tau)$ where the opacity, caused by neutral hydrogen, helium, and once ionized helium, stems from high lying cool gas in part carried aloft by flux emergence. The largest vertical field strengths in the photosphere are of order 2000~G or higher, while the mean strength is of order 80~G. At the height of \fex\ emission, 1.5~Mm above the photosphere and higher, the maximum (total) field strengths have fallen to roughly 300~G, which is towards the low end of the field strengths possible to measure with the method described here. While we believe that this is a fairly typical coronal field strength for an emerging ephemeral region, it is towards the low end of what is possible to measure with the technique described here, as evidenced by Figure~\ref{fig:gbtne}. We will therefore artificially increase by six times this field strength in much of the analysis that follows.

\section{Results} \label{sec:res}

\subsection{Intensities, MIT and absorption}\label{sec:int}

We have computed the synthetic intensities of several \fex\ lines from the two simulation snapshots considered. As mentioned above, we will focus on the flux emergence simulation in the main text while the equivalent figures for the MURaM simulation can be found in the appendix \ref{sec:appendix}. Figure~\ref{fig:inta} shows the \fex~174, 257.259, 257.261~\AA\  and 345 intensities. The 257.261 line has been computed assuming B=0 in panels c, h, and m, while in panels d, i, and n, we set a field strength six times larger than that found in the model to enhance the effects of the MIT. In the middle row, we also consider absorption. The intensities of the various lines differ in some cases by more than one order of magnitude. The \fex~174~\AA\ line is the strongest emitter. Note also that, for the high density model presented here, which has flux emergence, the \fex~257.261~\AA\ intensity is lower than 257.259~\AA\ (see also panels b and c in Fig.~\ref{fig:gtne}). For solar conditions at lower coronal densities the relative contributions of these two lines may be significantly different. A close look at the third and fourth columns of Figure~\ref{fig:inta}, shows that the 257.261 line is brighter in some locations if the MIT is included. However, the absorption also leads to significant changes in intensity for all lines in varying degrees. 

\begin{figure*}
    \includegraphics[width=0.99\textwidth]{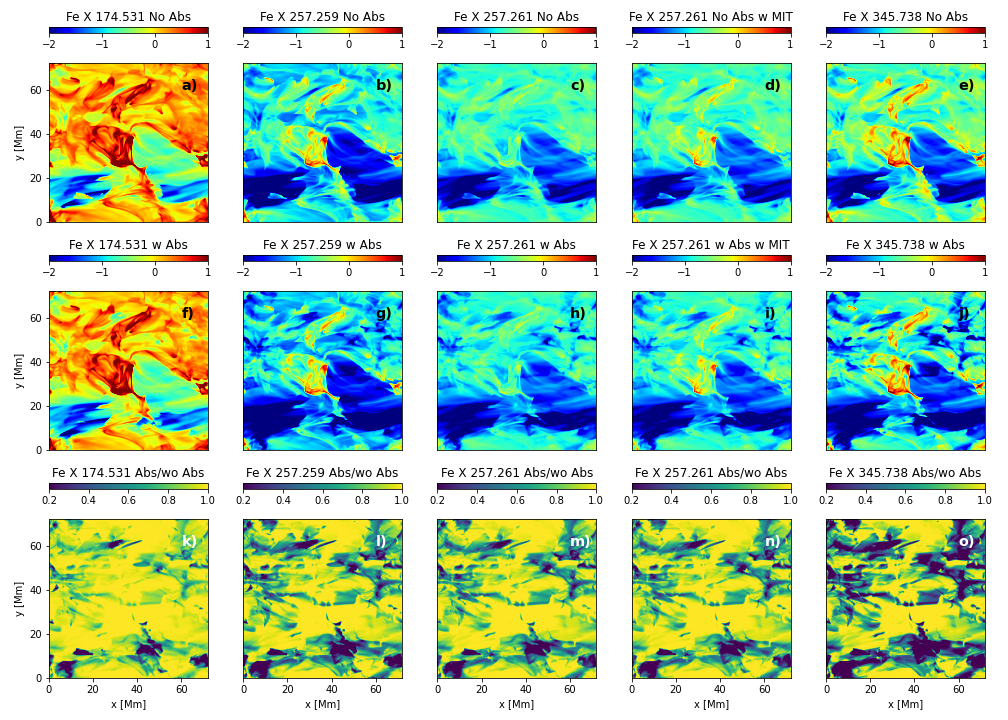}
	\caption{\label{fig:inta} Logarithmic scale intensities of \fex\ lines in W s$^{-1}$ cm$^-2$ sr$^{-1}$ without taking into account absorption shown in the top row, and intensities where absorption from high lying cool gas is included in the middle row. In the bottom row we show the ratio of the intensity of \fex\ lines computed taking into account bound-free absorption with respect to the intensity of these lines computed without taking into account absorption. The third and fourth columns show the difference between ignoring and including the MIT effect. From left to right we find the lines \fex~174, 257.259, 257.262, 345~\AA. See the corresponding figure for the AR simulation in the appendix (Fig.~\ref{fig:inta_hgcr}).}
\end{figure*}

To clarify the role of absorption from neutral cold gas, we compute the ratio of the line intensities with and without including absorption, i.e. we present the ratio of the top and middle rows of Figure~\ref{fig:inta}, shown in bottom row. As expected, the longer the wavelength, the greater the difference between the intensity with and without absorption. This dependence with wavelength comes about because the opacity scales with $\lambda^3$ (see Section~\ref{sec:syn}). It is also important to note that the opacity has contributions from neutral and singly ionized helium for spectral lines below 504~\AA\ and 228~\AA, respectively \citep{Anzer:2005ye,Rumph:1994qo}. This simulation, which has a significant amount of cool gas carried aloft as a result emerging flux, has large opacities and hence absorption in all lines for extended regions. Still, other regions have very little or zero absorption. A central question is thus if it is possible to discern which intensity variations are due to density, temperature, or magnetic field vs. which are due to absorption, or at least to find in what regions absorption is important. 

\begin{figure*}
    \includegraphics[width=0.99\textwidth]{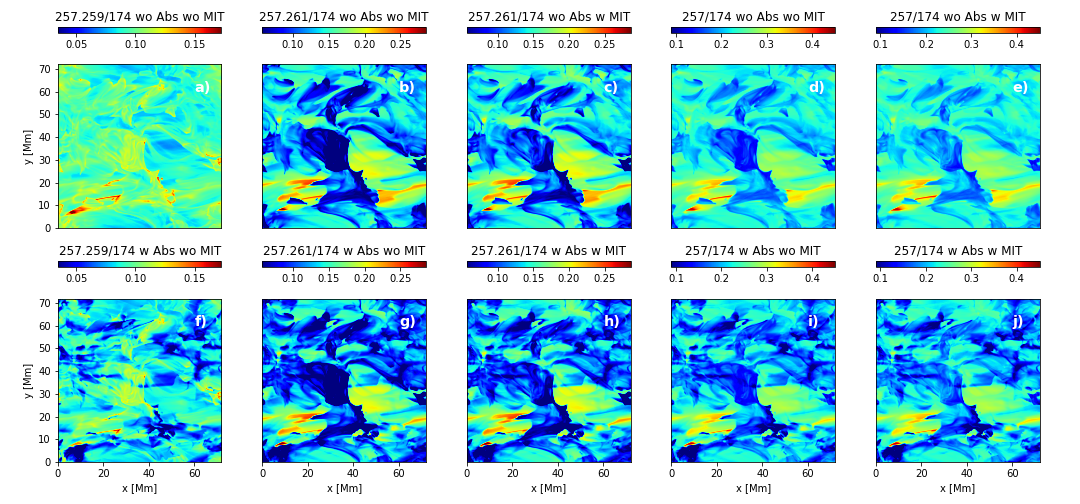}
	\caption{\label{fig:ratlines} The ratio between different spectral lines shows significant variations under different conditions, i.e., at different temperatures and densities and with and without absorption and MIT. The top row is without absorption and the bottom with absorption. The left two panels are \fex~257.259/\fex~174. Panels b, c, g, and h are for \fex~257.261/\fex~174 where panels c and h are with MIT. The right four panels are the blended case. See the corresponding figure for the AR simulation in the appendix (Fig.~\ref{fig:ratlines_hgcr}).}
\end{figure*}

Existing methods to derive the coronal magnetic field rely on the comparison of different spectral lines with varying sensitivity to the density and temperature to obtain an estimate of these thermodynamic parameters in the emitting plasma, \citep{Landi:2020ApJ...904...87L,Chen:2021ApJ...918L..13C,Brooks:2021ApJ...915L..24B} and thereafter to use these values when computing the magnetic field. To illustrate the complexity of this task, we compare \fex~257 lines with the 174~\AA\ line (Fig.~\ref{fig:ratlines}). The variation of the ratio of \fex~257.259 and 174~\AA\ (panel a) is due to different density and temperature sensitivity of the contribution functions for both lines. Absorption by cool gas also impacts the ratio of these two lines (panel f). The \fex~257.261/174 line ratio (panel b) reveals a much smaller temperature and density variation since the dependence of the \fex~257.261 and 174 lines is very similar (Fig.~\ref{fig:gtne}). However, the MIT effect (panel c) is relatively small, so it is critical to estimate the densities correctly in order to measure the magnetic field accurately. Unfortunately, absorption can play a large role when comparing \fex~257.261 and 174 (panels g and e). With the densities found in this model, the dominant contribution from \fex~257 comes from the \fex~257.259 line, which highlights that the density sensitivity and absorption by cool gas play a large role when comparing these two lines. We also notice that the absorption will change the height from which the emission predominantly originates.  This has an impact on the role of the MIT effect since different heights will have different magnetic field. In the following we present an inversion technique that may help us determine whether and where the data is not affected by absorption and retrieval of the magnetic field is possible (Section~\ref{sec:demb}).

\subsection{Inversions: DEM$(T, n_e, B)$ }~\label{sec:demb}

Based on the same idea described in  \citet{Cheung:2019ApJ...882...13C} we suggest a new approach to derive the magnetic field by using the multi-dimensional sensitivity (temperature, density, and magnetic field) of the spectral lines. The idea is to invert  the differential emission measure (DEM) from the observations which allows to constrain the temperature, densities and magnetic field strength. However, in contrast to previous efforts, we will take into account in the contribution function both the density and magnetic field dependence ($G_\lambda(T, n_e, B)$). In this case, the spectral line intensity ($I[\lambda]$) can be defined as follows: 

\begin{eqnarray}
I[\lambda] = \int_{T_0}^{T_1} \int_{n_{e0}}^{n_{e1}} \int_{B_0}^{B_1} G_\lambda\, \text{DEM}(T, n_e, B)\, dT\, dn_e\, dB 
\end{eqnarray}

\noindent where $\lambda$ is the wavelength of the spectral line, and the DEM$(T, n_e, B)= \int n_e\, n_H\, dl$ is the differential emission measure for each temperature, number density, and magnetic field value. The aim then becomes to invert the DEM given a set of spectral lines $I[\lambda]$ in order to find a unique solution $T$, $n_e$, and $B$ \citep[e.g.,][]{Cheung:2015ApJ...807..143C}. Ideally, one would also like to determine the amount of absorption as part of the inversion approach. However, this would add significant complexity in addition to expanding the number of free parameters in the inversion, thus complicating the convergence to the correct solutions. This is beyond the scope of the current work, however we will demonstrate that our approach can  identify regions in which there is significant bound-free absorption. 

We will use the same strategy as in  \citet{Cheung:2019ApJ...882...13C,Winebarger:2019ApJ...882...12W,DePontieu:2020ApJ...888....3D}, which solves the linear system with sparse solutions coming from the Lasso method \citep{tsinganos1980}. Lasso is implemented in the Python scikit learn module \citep{Pedregosa:scikit-learn}. See \citet{Cheung:2019ApJ...882...13C} for further details. 

For the inversion, we limit the temperature range to $\log{(T (K))} = [5.7, 6.3]$ to reduce the number of possible solutions as well as to enable reasonable array sizes. Note that the DEM will have five dimensions, i.e., two spatial dimensions in addition to $(T, n_e, B)$. We found that including \fex~174, 175, 184, 257 (blended), 345~\AA\ lines provides reasonable results also thanks to the different temperature sensitivity of these lines, whose excitation threshold is very different. Finally, this technique allows adjusting the weight of the spectral lines of interest. In the present case, we have increased by ten the weight accorded to the blended 257 lines (and thus the contribution function, accordingly). This enforces that the best fits will be found for the \fex~257 lines. Note that this line is the only one with potentially large magnetic field sensitivity. Finally, this inversion is controlled with a hyperparameter ($\alpha$) that controls the sparsity or entropy of the solutions. The greater the value of $\alpha$, the smaller the L1 norm and thus the sparsity of the solution \citep{Tibshirani:lasso,Pedregosa:scikit-learn}. In our case, we use a value that is as low as possible while avoiding artifacts from overfitting: $\alpha \sim 10^{-2}$. 

\begin{figure*}
    \includegraphics[width=0.7\textwidth]{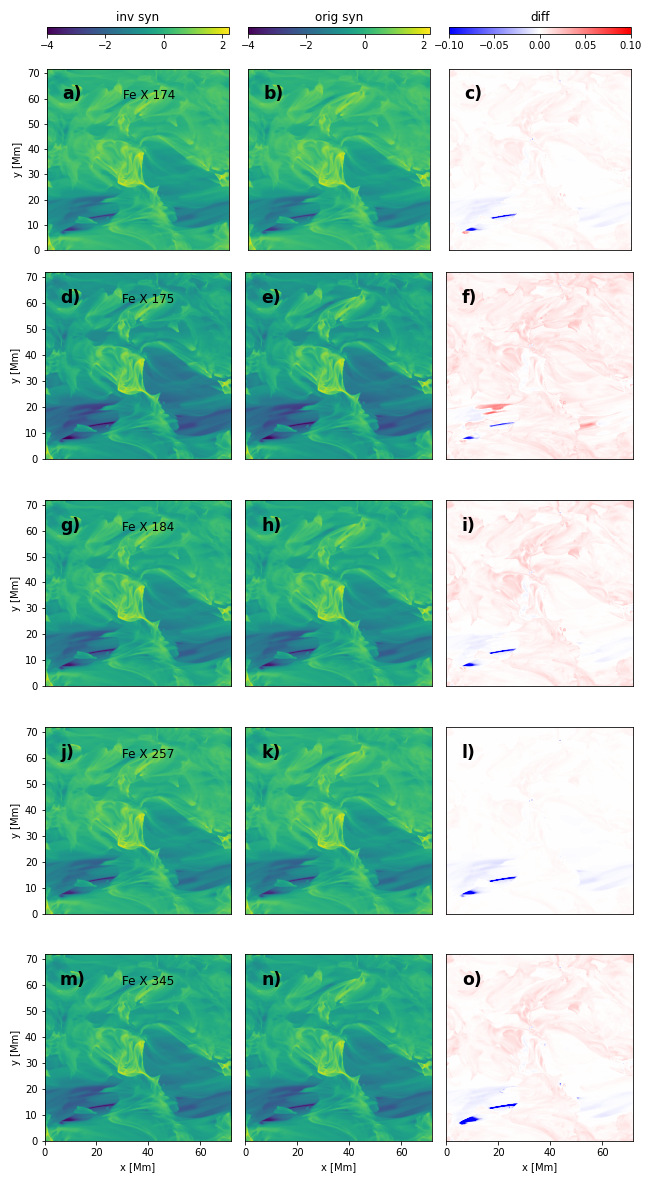}
	\caption{\label{fig:inv_syn} Comparison of the forward synthesis from the inverted DEM$(T, n_e, B)$ with the ground truth intensities. This comparison is without taking into account bound-free absorption. The left column shows the synthesis from the inverted DEM$(T, n_e, B)$, while the central column shows the logarithmic intensities calculated from the simulation. Finally, the right column shows the relative difference between intensities ($(I_{inv}-I_{syn})/I_{syn}$). From top to bottom we plot the \fex~174, 175, 184, 257 (blended), 334~\AA\ lines. See the corresponding figure for the AR simulation in the appendix (Fig.~\ref{fig:inv_syn_hgcr}).}
\end{figure*}

A nice feature of this technique is that one can form an  estimate of how good the solutions are by comparing the real intensities and the synthesized intensities from the inverted DEM. The case without absorption is shown in Fig.~\ref{fig:inv_syn}. Since we have placed a greater weight on \fex~257, the best fit is found for that line (panel l), whereas the spectral lines with artificially smaller weighted intensities (due to lower weights in the inversion) have larger errors. The right column of Fig.~\ref{fig:inv_syn} is the relative difference between the two left columns ($(I_{inv}-I_{syn})/I_{syn}$) and indicates where the derived magnetic field may not be accurate (see below). We note that the Lasso method gives higher weight to fits with sparse solutions that have a minimum DEM. This results in slightly lower intensities in general and therefore the right column is slightly red. 

\begin{figure*}
    \includegraphics[width=0.9\textwidth]{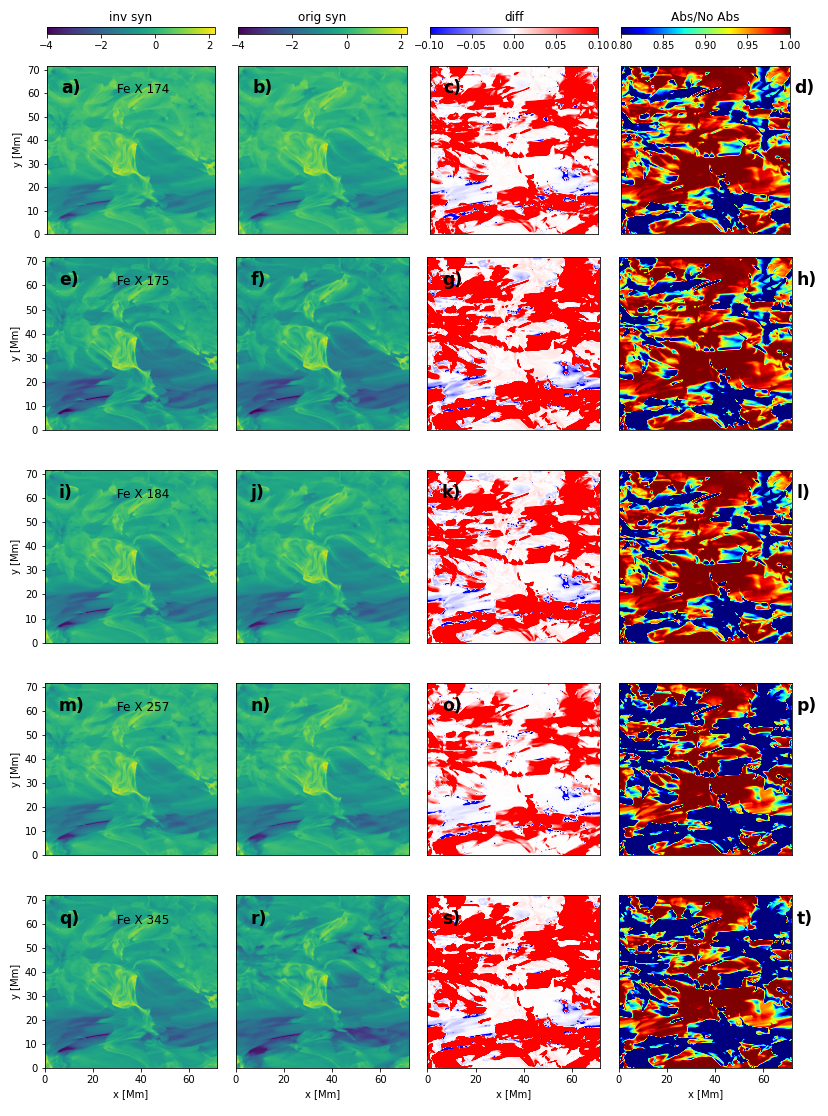}
	\caption{\label{fig:inv_synabs} This figure is equivalent to Fig.~\ref{fig:inv_syn}, but now with absorption. The right column is the ratio of the intensity with and without absorption. See the corresponding figure for the AR simulation in the appendix (Fig.~\ref{fig:inv_synabs_hgcr}).}
\end{figure*}

With the absorption included, the inversion does not find good solutions in those locations where absorption is important. We have used five spectral lines to derive the magnetic field, density, temperature and discern where the absorption may cause problems in doing so. In Fig.~\ref{fig:inv_synabs}, the same layout as in Fig.~\ref{fig:inv_syn} is used, but now for the case where absorption by cool gas is included. The last column shows the ratio between the intensity of the corresponding spectral line with and without absorption included. The third column shows an extremely good match with the right panel of the spectral line with the strongest absorption, as seen in panel t. So, the inversion is not finding good solutions in the DEM space to match the intensities for all spectral lines. The resulting synthesis from the inverted DEM does not match the original synthesis of the spectral lines. The largest absorption is for the \fex~345~\AA\ line, and this is why the errors match panel t. 

\begin{figure*}
    \includegraphics[width=0.95\textwidth]{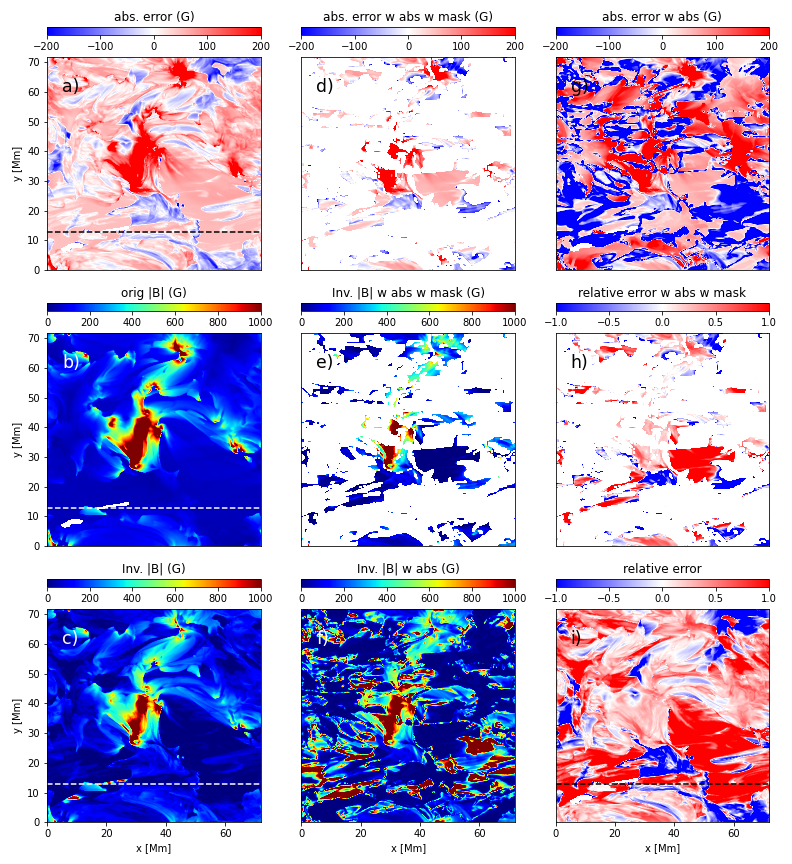}
	\caption{\label{fig:mag_maps} The inverted magnetic field ($B_i$) without absorption (panel c) and with absorption accounted for (panel f). The panel c shows the magnetic field from the simulation weighted with the emission (Eq.~\ref{eq:emmag}). The panel e shows the same as panel f, but with areas identified as having artifacts stemming from absorption masked out using the inversion method described in the text.  The top panels show the absolute error between panels b and c (in panel a), and panels b and e (in panel d) and panels b and f in panel g. The relative errors are shown in panels h (ratio between panel b and e) and i (ratio between panel b and f). The horizontal line is the cut shown in Figure~\ref{fig:mag_mult_comp}. See the corresponding figure for the AR simulation in the appendix (Fig.~\ref{fig:mag_maps_hgcr}).}
\end{figure*}

From the DEM$(T, n_e, B)$ we estimate the  average magnetic field along the LOS by weighting with the DEM and the source function ($G_{\lambda}(T, n_e, B)$) for each temperature and density bin as follows: 

\begin{eqnarray}
B_{i} = \frac {\int_{T_0}^{T_1} \int_{n_{e0}}^{n_{e1}} \int_{B_0}^{B_1} B\, \text{DEM}\, G_{\lambda}\, dT\, dn_e\, dB }{\int_{T_0}^{T_1} \int_{n_{e0}}^{n_{e1}} \int_{B_0}^{B_1} \text{DEM}\, G_{\lambda}\,  dT\, dn_e\, dB }
\label{eq:modb_mit}
\end{eqnarray}

\noindent note this could, similarly, be done for density or temperature diagnostics. To compare with the simulation, we compute the magnetic field in the model as follows:

\begin{eqnarray}
\text{EM}_{s} = n_{e_s}\, n_{H_s}\, G_{\lambda}(T_{s},n_{e_s},|B_{s}|); \\ 
B_{s} = \frac {\int B_{s}\, \text{EM}_{s} dl }{\int \text{EM}_{s} dl}~\label{eq:emmag}
\end{eqnarray}

\noindent where the subscript $s$ refers to the values from the simulation and $i$ refers to the inversions. The resulting maps of the inverted magnetic field ($B_i$) without (panel c) and with absorption (panels e and f), and the simulation in the numerical model ($B_s$) (panel b) are shown in Figure~\ref{fig:mag_maps}. The absolute error is shown without and with absorption in the top row (panel d includes the mask from panel s in Figure~\ref{fig:inv_synabs}) and the relative error is shown panels h (with mask and absorption) and i (without absorption). Without absorption (panel c), most of the magnetic field features are well resolved. However, the strongest magnetic field strengths in the inverted data seem to underestimate those found in the model. In contrast, the inverted magnetic field with absorption is filled with artifacts due to this absorption (panel f). Fortunately, our method allows the identification of such regions and thereby the possibility of masking out regions with large impact from absorbing gas (panel e, d, and h) and thereby to select the areas clean from absorption. The absolute error indicates that the high values of magnetic field are underestimated by up to a few hundred Gauss (red) and some weak values are overestimated by up to a several tens of Gauss (blue). The relative error panel shows that this error becomes smaller for stronger magnetic fields. See also below for more details.

\begin{figure*}
    \includegraphics[width=0.9\textwidth]{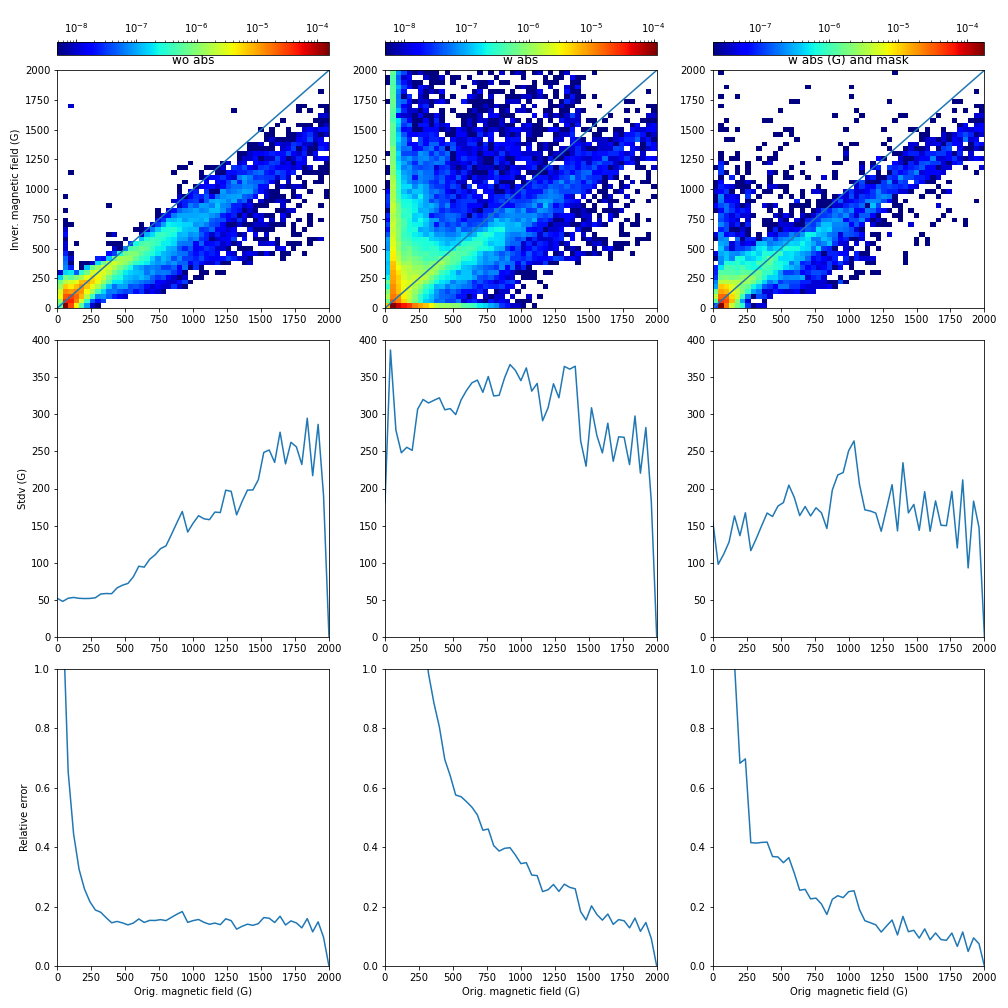}
	\caption{\label{fig:mag_hist} 2D histograms of the inverted magnetic field ($B_i$) without absorption with the magnetic field from the simulation ($B_s$) shows relatively good correlation (top-left panel). However, the correlation is bad for those pixels with absorption (top-middle panel). The mask from the inversion technique (see main text in Section~\ref{sec:demb} and Figure~\ref{fig:mag_maps}) allows to removes the bad pixels (top-right panel). The absolute standard deviation and relative error as a function of magnetic field are shown in the corresponding middle and bottom panels. See the corresponding figure for the AR simulation in the appendix (Fig.~\ref{fig:mag_hist_hgcr}).}
\end{figure*}

Indeed, a set of 2D histograms reveals the quality of the correlation between $B_s$ and $B_i$. Figure~\ref{fig:mag_hist} shows that $B_i$ is underestimated at high values. \citet{Chen:2021ApJ...918L..13C} suggested that their under (or over) estimates may come from uncertainties in the temperature diagnostic. Most likely this is not the case for our study since our method also takes into account temperature variations. Another possible scenario could explain the under-estimated values: As mentioned above, the Lasso method gives higher weight to fits with sparse solutions that have a minimum DEM. Those solutions tend to be biased to lower magnetic strengths. We see the same trend for the density parameter, but not for the temperature (not shown here). If this error is systematic across many realizations, one may be able to correct for such underestimates (see also appendix). This might depend on the lines selected for inversion. Further investigation is needed to determine whether such a systematic correction is possible. The 2D histograms and the absolute standard deviation (middle row) as a function of magnetic field shows that the deviation can be from a few tens of Gauss at low magnetic field values to up to 200 G for regions without absorption. The relative error is large (above 40\%) for magnetic field values lower than 250~G and only above those values the relative error decreases to 18\%. Regions with significant absorption experience larger errors. 

\begin{figure}
    \includegraphics[width=0.49\textwidth]{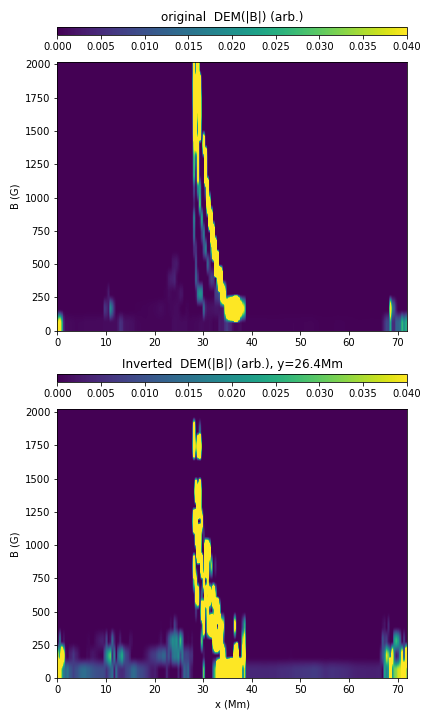}
	\caption{\label{fig:mag_mult_comp} The inverted DEM$(B)$ could potentially reveal the presence of multi-magnetic field components as shown for a specific horizontal dashed line in top row of Fig.~\ref{fig:mag_maps}. Top panel corresponds to the DEM$(B)$ from the simulation and the bottom panel for the inverted DEM$(B)$ where we integrated temperature and density space. See the corresponding figure for the AR simulation in the appendix (Fig.~\ref{fig:mag_mult_comp_hgcr}). The associated online animation scans along the y-axis.}
\end{figure}

The inversion could allow an investigation of the DEM as a function of temperature, density, or, for this work, the magnetic field, in a similar fashion as is done when estimating the DEM as a function of temperature using AIA data. The temperature coverage of the selected lines we study here is of course very limited. In addition, while determining the density from DEM may work, but this extension is outside of the scope of this paper. Figure~\ref{fig:mag_mult_comp} shows that in some places, the DEM$(B)$, i.e. integrated with temperature and density, can identify some multi-magnetic field components which correspond to different structures along the LOS, e.g., around $x=[30,35]$~Mm. In this figure, we degraded the magnetic field bin to 50~G since the standard deviation shown in Figure~\ref{fig:mag_maps} has no better values than those. We do not expect to achieve better accuracy of the DEM(B) than for the full integration done in equation~\ref{eq:modb_mit}. While the values are slightly inaccurate, the inversion appears to be doing a good job in revealing multiple components in some locations and thus could provide, in principle, more observables than the least-square methods or line intensity ratios used in earlier studies. This could have the potential of revealing multi-magnetic field structures along the LOS. 

In summary, the bound-free absorption by cool plasma at coronal heights has been ignored in previous studies. We have investigated the MIT effect taking into account the absorption and provide a new technique to derive the magnetic field. We find that absorption can locally have a major impact on the determination of the magnetic field. Our new method allows to identify regions with significant absorption, providing more confidence to the derived magnetic field in absorption-free locations, using either line intensity ratios, the least-square methods or the one presented here. 

\section{Conclusions and discussion}\label{sec:con}

This study has utilized two 3D radiative MHD numerical models characterized by two very different magnetic field configurations. The first mimics an emerging ephemeral region using the Bifrost code (main text). The second is a simulated AR containing the early stages of new flux emergence into the existing active region configuration (Appendix~\ref{sec:appendix}). We synthesize the emission of highly ionized iron atoms in the EUV to further understand the potential for diagnostics from the magnetically sensitive \fex~257~\AA\ line. In total we have synthesized \fex~174, 175, 177, 184, 257 (the two blended lines), and 345~\AA\, taking into account their temperature, density and magnetic dependence. In addition to modeling the effects of the local state of the emitting plasma, we consider the possible effects of the presence of significant amounts of cool gas lying above the emitting region and synthesize spectra according to two different scenarios: with and without including bound-free absorption by neutral hydrogen and helium, and single ionized helium. 

Our analysis of the synthesized spectral lines reveals that the absorption can cause considerable difficulties in estimating both the density and the magnetic field from the various lines. Furthermore, for the two models, we found out that the relative intensity variation of \fex~257~\AA\ in comparison to \fex~174 due to the magnetic field is comparable to or even less important than the variation due to absorption of cold plasma in those locations with overlaying cold plasma. 

In the second part of this study, we describe a novel inversion technique to derive the magnetic field, density and temperature. This inversion technique has earlier been used to derive the DEM; for instance from AIA observations \citep{Cheung:2015ApJ...807..143C}, or to disambiguate the spectrum from multi-slit observations or slitless observations, e.g., MUSE \citep{Cheung:2019ApJ...882...13C} and COSIE \citep{Winebarger:2019ApJ...882...12W}. The novelty in this study is that we consider the temperature, density, and for \fex~257~\AA\ also the magnetic field dependence of the contribution function. So, in principle, this approach could retrieve the magnetic field as well as the temperature and densities. Another interesting diagnostic that this tool may be able to provide is the product of the filling factor with the column depth for each temperature, magnetic field, and/or density bin within the FOV. This is possible because the inversion has derived the DEM and the number density for each bin. This aspect requires further investigation in future work.

When absorption is not included in the analysis, our results show similar results as those reported by \cite{Chen:2021ApJ...918L..13C}. We estimate that the accuracy we can achieve with this method is 50-200~G in locations without significant absorption. The absolute accuracy is highest at lower magnetic field values and rises to 200~G at high magnetic field values. Consequently, we obtain reasonable relative errors above 250~G. We also noticed that the MURAM simulation has a slower improvement in the relative error with field strength between 250-600~G. These slightly worse errors seem to come from regions with an overestimate of the magnetic field. The LOS analysis of the DEM(B) seems to indicate that these regions have more than one component of the magnetic field, and one of them is nearly zero Gauss. These estimates are less optimistic than those provided by \citet{Landi:2020ApJ...904...87L}, who carried out an uncertainty analysis and determined that the minimum detectable magnetic field in active regions was of the order of 50~G. Unfortunately, observations cannot be compared with actual values as we did with the numerical models. We also noted that we used a different line list. It is unclear what typical magnetic field values in the corona are. Based on the simulations we have used here, only a modest fraction of the field of view shows magnetic field values stronger than this sensitivity limit. In fact, for the flux emergence simulation we had to increase the coronal magnetic by six in order to obtain a measurable range of field values. Note that this may limit the applicability of the MIT method, either using the current method or the least-square or line ratio methods, to regions with strong coronal magnetic field such as footpoints, lower legs of AR loops, loops associated with sunspots, etc.  

The least-square method used by previous studies \citep[e.g.,][]{Chen:2021ApJ...918L..13C} also underestimates the magnetic field for a specific combination of spectral lines. We noticed that in our method the inversion can provide better results by varying the weights on the intensities. In this way we can control the inversion to better fit the lines of interest. In this work, we have found the desired weights with trial and error. However, the method to find the best weights probably can be automated and fine-tuned. This is something that can be investigated in the future.  Another possibility of improving the inversions would be by looping the inversions, by first starting with low resolution and large ranges for temperature, density, and magnetic field. Then, the output of this first loop can be used to trim the temperature, density, and magnetic field range and thereby reduce their bin sizes, thus improving accuracy. Finally, the sparsity of the solution can be controlled with the hyperparameter $\alpha$. This parameter may need to be adjusted while using this method for different observations. Such a detailed study will be the subject of a follow-up paper.

This technique provides estimates that are similar to \cite{Chen:2021ApJ...918L..13C} with the addition of being able to mask regions with absorption and with the potential of discerning multi-magnetic field components. In addition, our method self-consistently and simultaneously solve for the density, temperature and magnetic field sensitivity of these spectral lines. We note that the Lasso method gives higher weight to fits with sparse solutions that have a minimum DEM: This results in an underestimate of the density and magnetic field for large values of the density and magnetic field, respectively. This error appears systematic and could probably be corrected, but further studies are necessary and we expect that the accuracy of our result will also depend on the spectral lines available for inversion.  Similar errors have been found in previous studies using least squared methods but the systematic error in those cases may come from a different reason \cite[see for details][]{Chen:2021ApJ...918L..13C}. In principle, one could expand the parameter range of the inversion and include velocities as has been outlined for the disambiguation of MUSE spectra \citep{Cheung:2019ApJ...882...13C,DePontieu:2020ApJ...888....3D}. This would allow one to further isolate structures along the LOS, with temperature, density and magnetic field bins as well as velocity bins. The NLFFF method provides similar error estimates \citep[of the order of 20\%, e.g.,][]{Metcalf:2008bd,De-Rosa:2009rq}. So for field strongest to 250~G the MIT method is a good complement on constraining the magnetic field. 

The main advantage of our method is that it provides a powerful tool to mask regions that may suffer absorption. The reason is that once there is absorption, the inversion will not find reasonable solutions. In these locations the comparison between the observation and prediction from the forward synthesis with large absorption do not match. Similarly, one would expect that this sort of mismatch could also be used for regions suffering significant non-equilibrium effects or instrumental artifacts, e.g. discrepancies in the absolute calibration. In order to expand the parameter range of the analysis on the spectral line properties and methods we used two very different simulations. One with an active emerging flux with relatively large amount of absorption. We found that for this scenario, the coronal magnetic field is too weak and we had to increase by 6 times to get enough signal on the MIT effect. The second one is with lower spatial resolution, less absorption and stronger magnetic field. We notice that for both regions the systematic error in the derived magnetic field (under-estimate of the magnetic field) is the same giving support to the possibility that this can be corrected. 

This method would be of great interest for Hinode/EIS observations. However, Hinode/EIS does not include \fex~345. Still, obtaining estimates of the coronal magnetic field strength should be possible as already shown by \cite{Brooks:2021ApJ...915L..24B} using line intensity ratios. The method will also be of even greater value in measuring coronal magnetic fields using upcoming EUVST observations which can complement with relevant iron lines to further constrain the DEMs.

\appendix

\section{radiative transfer}~\label{sec:appendix_rad}

The contribution function ($G_\lambda (T, n_e, B)$) used in eq.~\ref{eq:int} is defined as

\begin{equation}
G_\lambda (T, n_e, B) = \frac{N_j{\left({X^{+m}}\right)}}{N{\left({X^{+m}}\right)}}\frac{N{\left({X^{+m}}\right)}}{N{\left({X}\right)}}\frac{N{\left({X}\right)}}{N{\left({H}\right)}}\frac{N{\left({H}\right)}}{n_e}\frac{A_{ji}}{n_e}h\nu_{ji}
\end{equation}

\noindent
where $N_j{\left({X^{+m}}\right)}/N{\left({X^{+m}}\right)}$ is the upper level population which can depend on electron density and temperature and, in the case of the level giving rise to the \fex~257.261~\AA\ line, also on the magnetic field strength $B$ as described by \citet{Li:2015ApJ...807...69L,Li:2016ApJ...826..219L}. $N{\left({X^{+m}}\right)}/N{\left({X}\right)}$ is the ion fraction (strongly dependent on the electron temperature), $N{\left({X}\right)}$ is the element abundance, $N{\left({H}\right)}/n_e$ is the ratio between total H density and free electron density, and $A_{ji}$ is the Einstein coefficient for spontaneous emission for the transition between upper level $j$ and lower level $i$, having a frequency $\nu_{ji}$ \citep{Phillips:2008uxss.book.....P}.

The bound free absorption used in this work is by nuetral hydrogen, leium as well as singly ionized helium following the recipes of \citet{Anzer:2005ye}. In short, 

\begin{eqnarray}
\tau & = & n_H \{ (1-F_{HI})\sigma_H + \nonumber \\
 & & A_{He} [(1-F_{HeI}-F_{HeII})\sigma_{HeI} +  F_{HeI}\sigma_{HeII}]\}
\end{eqnarray}

\noindent where $n_H$ is the number density of hydrogen, $A_{He}$ the helium abundance, and $F_HI$, $F_{HeI}$, and $F_{HeII}$ are the ionization fraction for hydrogen, and helium. The photoionization cross sections can be found in \citet{Mihalas:1978stat.book.....M} as follows: 

\begin{eqnarray}
\sigma_H(\lambda) = \sigma_0 g_H(\lambda) (\lambda/912)^3 \\ 
\sigma_{HeII}(\lambda) = 16\sigma_0 g_{HeII}(\lambda) (\lambda/912)^3 \\ 
\log_{10}{\sigma_{HeI}(\lambda)} = \sum_i^7 c_i \log_{10}(\lambda)^i 
\end{eqnarray}

\noindent see \citet{Anzer:2005ye} for further details, and we used the coefficients $c_i$ in \citet{Rumph:1994qo}. We would like to highlight here the dependence of the cross sections on $\lambda$ which provides a different absorption coefficient for each spectral line of varying wavelength. 

The CHIANTI v.10 model for \fex\ includes 552 fine structure levels, which allow users to account for the effects of cascades. Collisional excitation rates are taken from the R-Matrix calculations of \citet{DelZanna:2021ApJ...909...38D} while the Einstein coefficients come from \citet{Wang:2020ApJS..246....1W} for the $n=3$ levels and \citet{DelZanna:2021ApJ...909...38D} for $n=4$ levels. In general, the accuracy of predicted contribution functions depends critically on the accuracy of two sets of data: 1) the ionization and recombination data used to calculate ion fractions, and 2) the atomic data, transition rates, and electron-ion collisional excitation rates used to calculate level populations. In the case of \fex, the CHIANTI model has evolved considerably since the first release, and both sets of data have been improved multiple times. Level populations have changed by a factor of 2 from the first version of CHIANTI in 1997 to the latest one; improvements have been substantial, but the changes that occurred in the last three CHIANTI versions (V8, V9 and V10) provide negligible differences even if the value for the radiative transition probability of the E1 $3s^23p^5~^2P_{1.5} - 3s^23p^4(^3P)3d~^4D_{2.5}$ transition at 257.259~\AA\ has changed by a factor 2 between V8 and V10. The reason is that radiative decay through the E1 257.259~\AA\ line is the only decay avenue available for the $^4D_{2.5}$ level, so that the overall line intensity did not change -- thus, we expect no change in the E1 transition if the more recent decay rate of \citet{Li:2021ApJ...913..135L} is used in place of the \citet{Wang:2020ApJS..246....1W} value. The largest source of uncertainty at the moment are the ion fractions, as the maximum abundance temperature have shifted towards higher temperatures, resulting in temperature-dependent differences of up to a factor 3; however, this change does not affect the MIT technique, which is entirely based on intensity ratios from lines emitted by \fex\ only.

\section{Collection of figures for the MURaM AR simulation}~\label{sec:appendix}

In the main text we show a case where a significant amount of cold plasma is lifted to coronal heights due to flux emergence. The simulation may be similar in nature to an ephemeral region or a newly forming active region. Other regions may experience similar or even more absorption, e.g., as found in UV bursts, moss, coronal rain, or limb observations to cite a few. Here, we now consider a mature AR where cool loops have had time enough to drain and which is characterized by long hot loops. This simulation has been performed with the MURaM code. 

\subsection{Intensities, MIT and absorption}

\begin{figure*}
    \includegraphics[width=0.98\textwidth]{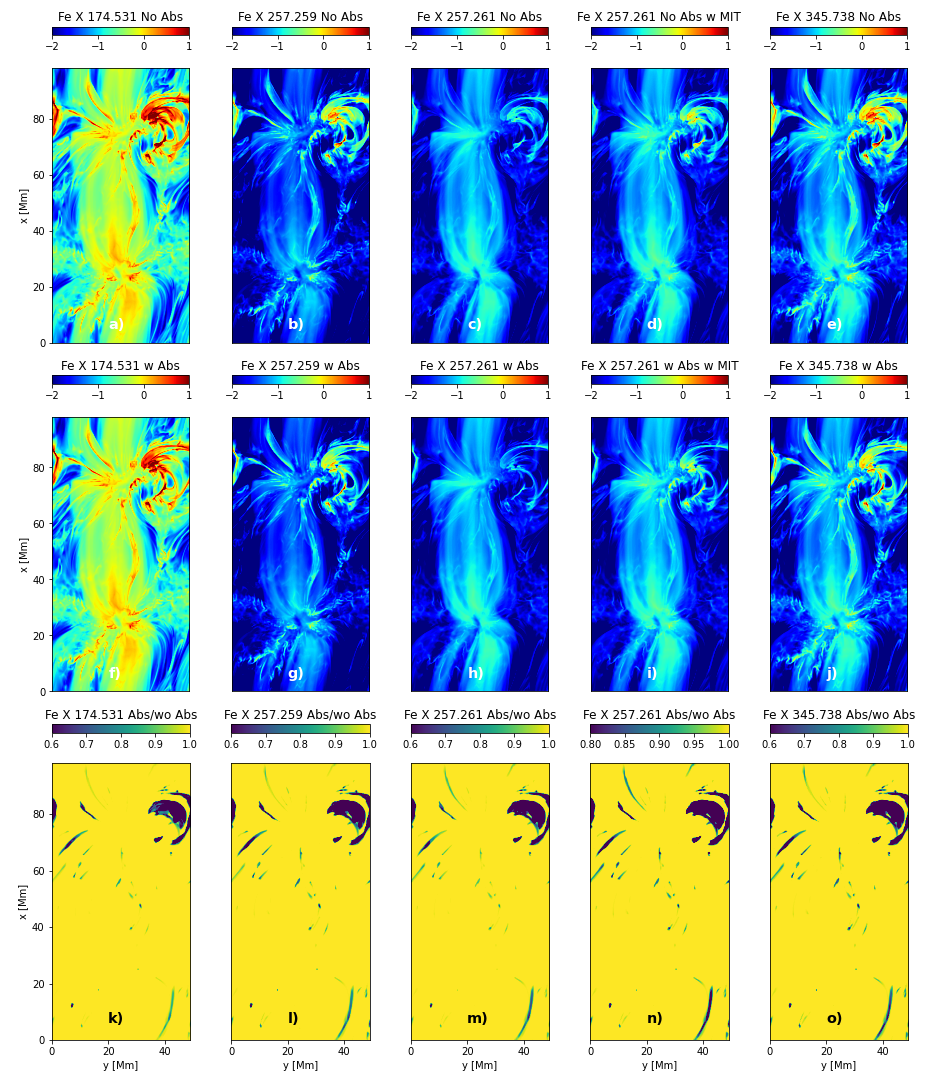}
	\caption{\label{fig:inta_hgcr} Equivalent to Figure~\ref{fig:inta} for the AR region model using MURaM. Logarithmic scale intensities of \fex\ lines with and without absorption shown in the top and middle row, respectively. The ratio of \fex\ lines computed taking into account absorption vs lines computed ignoring the effects of cool gas are shown in the bottom row. The third and fourth columns show the difference between ignoring and including the MIT effect. From left to right we find the lines \fex~174, 257.259, 257.262, 345~\AA.}
\end{figure*}

\begin{figure*}
    \includegraphics[width=0.98\textwidth]{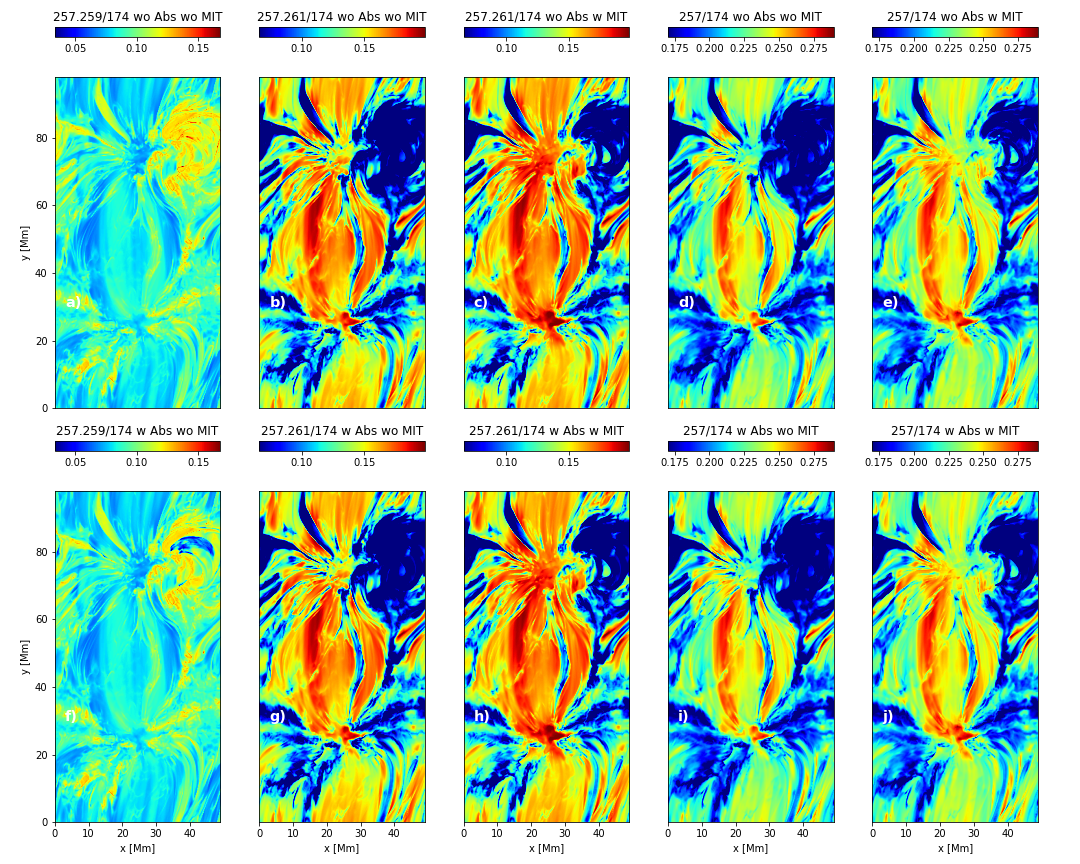}
	\caption{\label{fig:ratlines_hgcr} Equivalent to Figure~\ref{fig:ratlines} for the AR region model using MURaM. The ratio between different spectral lines. The top row is without absorption and the bottom with absorption. The left two panels are \fex~257.259/174. Panels b, c, g, and h are for \fex~257.261/174 where panels c and h are with MIT. The right four panels are the blended case.}
\end{figure*}

Like for the previous model, the comparison between the different spectral lines become a multi-dimensional problem that depends on temperature, density, magnetic field as well as absorption (Figures~\ref{fig:inta_hgcr}, and~\ref{fig:ratlines_hgcr}).  This model is not as dominated by flux emergence, but does include some localized flux emergence (top-right). It also shows some 1 MK emission in loop footpoints (``moss") with some absorption, but this is reduced compared to the extensive presence of moss in the solar atmosphere \citep{DePontieu:2009ApJ...702.1016D}. This may be because there are fewer chromospheric jets in this simulation because of its low spatial resolution. The jets that are present lift less cold plasma and thus cause less absorption at the footpoints of the hot loops (Fig~\ref{fig:inta_hgcr}). Large extended loops have negligible absorption, whereas low lying loops suffer from absorption. This model corresponds to an AR and has a magnetic field that is more typical for the corona, so we did not increase the magnetic field. 

\subsection{Inversions: DEM$(T, n_e, B)$}

\begin{figure*}
    \includegraphics[width=0.9\textwidth]{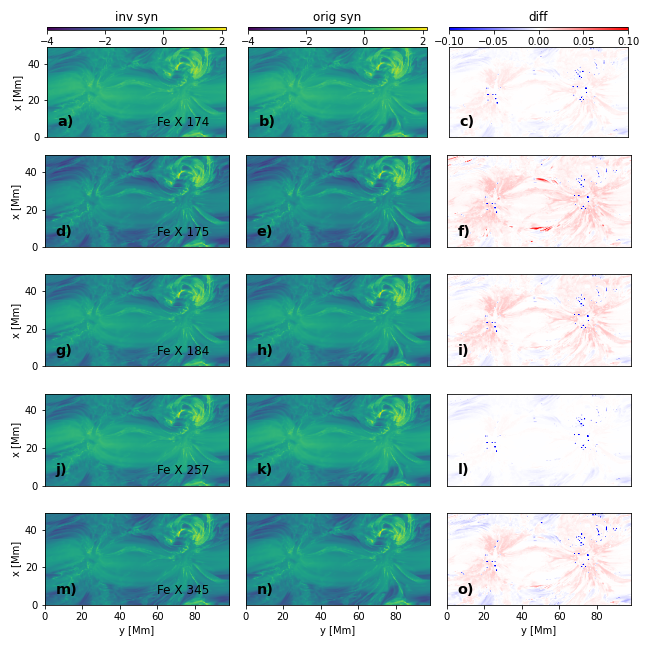}
	\caption{\label{fig:inv_syn_hgcr} Equivalent to Figure~\ref{fig:inv_syn} for the AR region model using MURaM. Comparison of the forward synthesis from the inverted DEM$(T, n_e, B)$ with the ground truth intensities. This comparison is without absorption. The left column shows the synthesis from the inverted DEM$(T, n_e, B)$, while the central column shows the logarithmic intensities calculated from the simulation. Finally, the right column shows the relative difference between intensities ($(I_{inv}-I_{syn})/I_{syn}$). From top to bottom we plot the \fex~174, 175, 184, 257 (blended), 334~\AA\ lines. }
\end{figure*}

\begin{figure*}
    \includegraphics[width=0.9\textwidth]{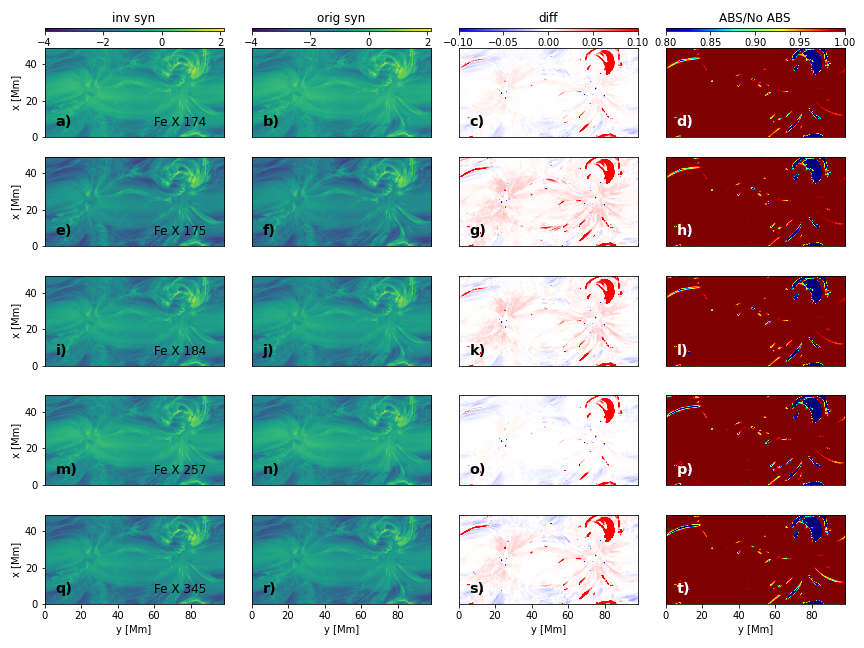}
	\caption{\label{fig:inv_synabs_hgcr} Equivalent to Figure~\ref{fig:inv_synabs} for the AR region model using MURaM. This figure is the same to Fig.~\ref{fig:inv_syn_hgcr}, but now with absorption. The right column is the ratio of the intensity with and without absorption.}
\end{figure*}

The inversion is doing a good job on reproducing the observables (Fig.~\ref{fig:inv_syn_hgcr}). Similarly this technique is able to reveal where the absorption is happening allowing to mask those locations from the inversion (Fig.~\ref{fig:inv_synabs_hgcr}). 

\begin{figure*}
    \includegraphics[width=0.95\textwidth]{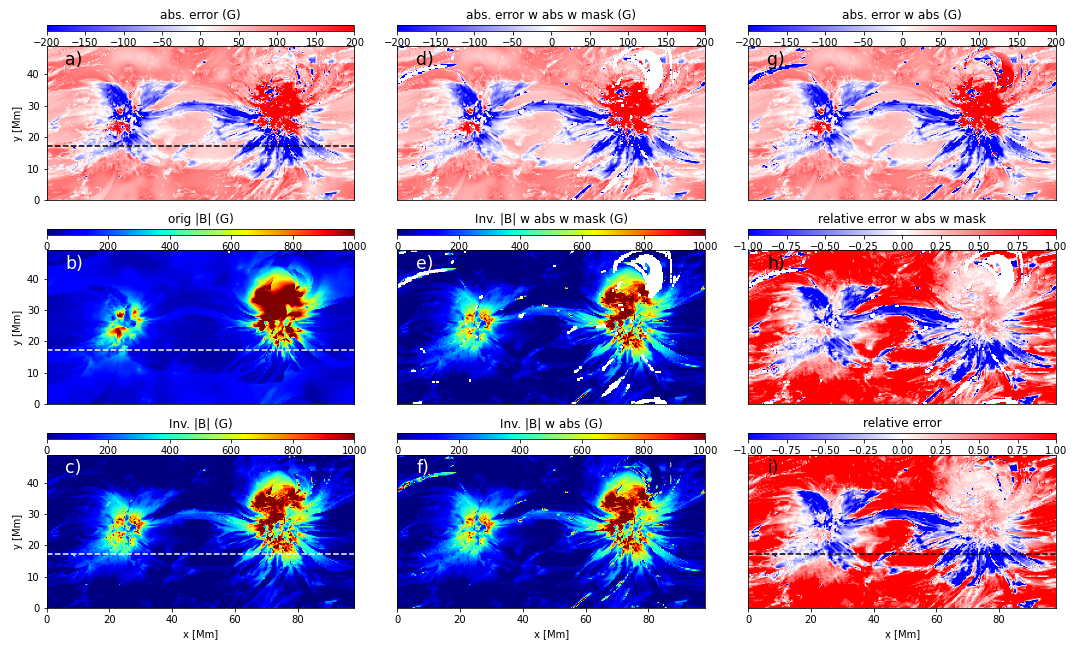}
	\caption{\label{fig:mag_maps_hgcr} Equivalent to Figure~\ref{fig:mag_maps} for the AR region model using MURaM. The inverted magnetic field ($B_i$) without absorption is shown in panel c. The same is shown with absorption accounted for  in panel f. Panel c shows the magnetic field from the simulation weighted with the emission (Eq.~\ref{eq:emmag}). Panel e shows the same as panel f, but with areas identified as having artifacts stemming from absorption masked out using the inversion method described in the text.  The top panels show the absolute error between panels b and c (in panel a), and panels b and e (in panel d) and panels b and f in panel g. Whereas the relative errors are shown in panels h (ratio between panel b and e) and i (ratio between panel b and f). The horizontal line is the cut shown in Figure~\ref{fig:mag_mult_comp_hgcr}.}
\end{figure*}

\begin{figure*}
    \includegraphics[width=0.9\textwidth]{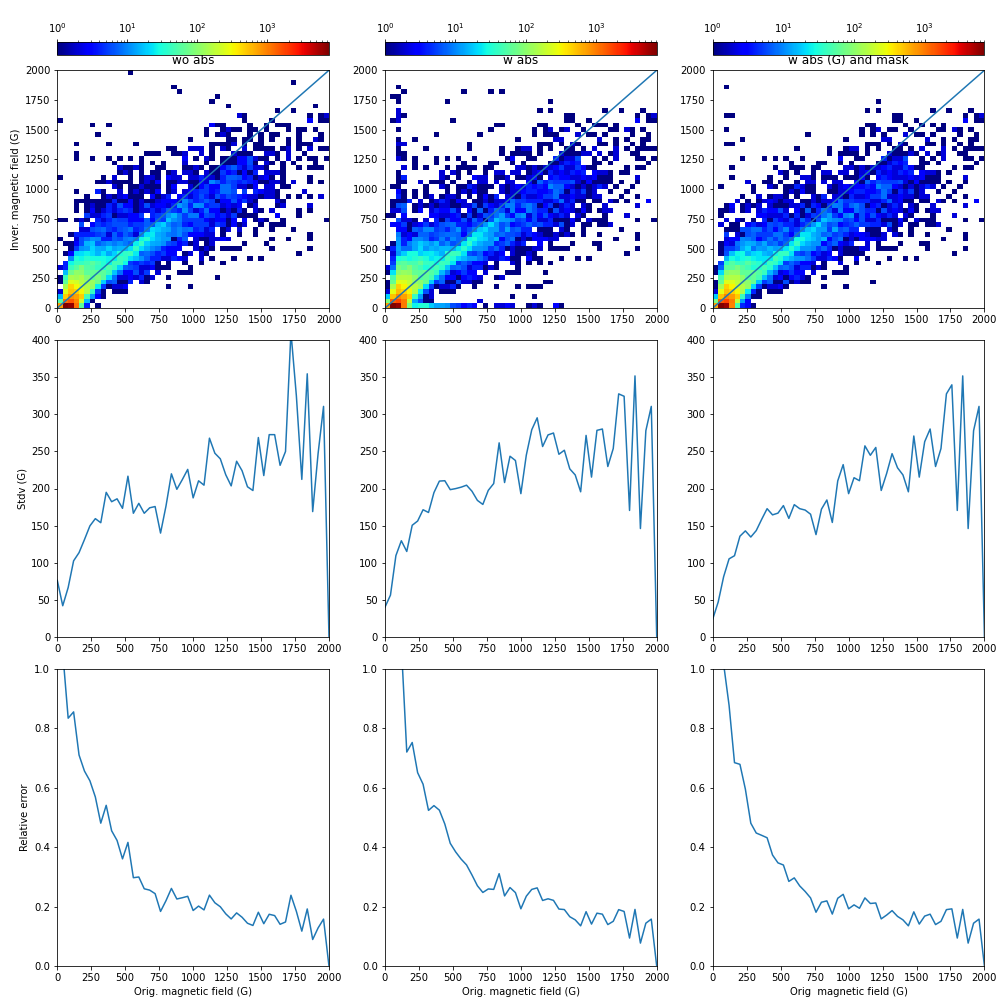}
	\caption{\label{fig:mag_hist_hgcr} Equivalent to Figure~\ref{fig:mag_hist} for the AR region model using MURaM. 2D histograms of the inverted magnetic field ($B_i$) without absorption with the magnetic field from the simulation ($B_s$) shows good correlation (top-left panel). However, the correlation gets bad for those pixels with absorption (top-middle panel). The mask from the inversion technique (see main text in section~\ref{sec:demb} and Figure~\ref{fig:mag_maps_hgcr}) allows to removes the bad pixels (top-right panel). The absolute standard deviation and relative error as a function of magnetic field are shown in the corresponding middle and bottom panels.}
\end{figure*}

\begin{figure*}
    \includegraphics[width=0.49\textwidth]{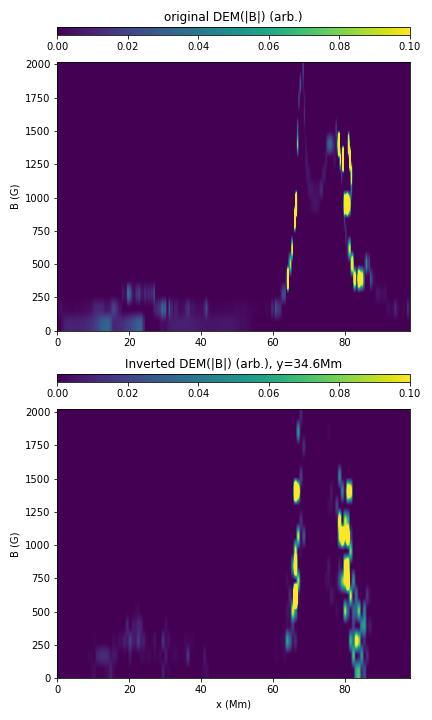}
	\caption{\label{fig:mag_mult_comp_hgcr}  Equivalent to Figure~\ref{fig:mag_mult_comp} for the AR region model using MURaM. Top panel corresponds to the DEM$(B)$ from the simulation and the bottom panel for the inverted DEM$(B)$ where we integrated temperature and density space for a specific horizontal dashed line in top row of Fig.~\ref{fig:mag_maps_hgcr}. The associated online animation scans along the y-axis.}
\end{figure*}

The resulting derived magnetic field shows results consistent with the ephemeral simulation from Bifrost (Fig.~\ref{fig:mag_maps_hgcr}, and~\ref{fig:mag_hist_hgcr}). Likewise, the derived magnetic field underestimates the real magnetic field, providing support to the possibility to apply a systematic correction. Note that, DEM analysis in Fig.~\ref{fig:mag_mult_comp_hgcr} reveals that it could be used to separate different magnetic field components along the LOS. 

It is also worth pointing out that in the MURaM simulation, we have some regions with an overestimate of the magnetic field strength ($[x,y]\approx[20,35]$~Mm). The animated Figure~\ref{fig:mag_mult_comp_hgcr} reveals that the overestimate may be because there are at least two magnetic field components, and one is very close to zero. The inversion provides a value between the two components of the magnetic field.

\acknowledgements{\longacknowledgment} 

\bibliographystyle{aasjournal}
\bibliography{aamnemonic,collectionbib}

\end{document}